\documentclass[fleqn,twoside]{article}
\usepackage{espcrc2}
\usepackage{latexsym}
\usepackage{euscript}

\usepackage{graphicx}
\usepackage[figuresright]{rotating}

\def\simge{\mathrel{%
   \rlap{\raise 0.511ex \hbox{$>$}}{\lower 0.511ex \hbox{$\sim$}}}}   
\def\simle{\mathrel{   
   \rlap{\raise 0.511ex \hbox{$<$}}{\lower 0.511ex \hbox{$\sim$}}}}   
\def\slashchar#1{\setbox0=hbox{$#1$}           
   \dimen0=\wd0                                 
   \setbox1=\hbox{/} \dimen1=\wd1               
   \ifdim\dimen0>\dimen1                        
      \rlap{\hbox to \dimen0{\hfil/\hfil}}      
      #1                                        
   \else                                        
      \rlap{\hbox to \dimen1{\hfil$#1$\hfil}}   
      /                                         
   \fi}                                         %

\def\simge{\mathrel{%
   \rlap{\raise 0.511ex \hbox{$>$}}{\lower 0.511ex \hbox{$\sim$}}}}   
\def\simle{\mathrel{   
   \rlap{\raise 0.511ex \hbox{$<$}}{\lower 0.511ex \hbox{$\sim$}}}}   
\def\slashchar#1{\setbox0=\hbox{$#1$}           
   \dimen0=\wd0                                 
   \setbox1=\hbox{/} \dimen1=\wd1               
   \ifdim\dimen0>\dimen1                        
      \rlap{\hbox to \dimen0{\hfil/\hfil}}      
      #1                                        
   \else                                        
      \rlap{\hbox to \dimen1{\hfil$#1$\hfil}}   
      /                                         
   \fi}

\newcommand{\AmS}{{\protect\the\textfont2
  A\kern-.1667em\lower.5ex\hbox{M}\kern-.125emS}}

\newcommand{\ba}{\begin{equation} \left\{ \begin{array}{lr}}
\newcommand{\ea}{\end{array} \right. \end{equation}}

\newcommand{\bea}{\begin{eqnarray}}
\newcommand{\eea}{\end{eqnarray}}
\newcommand{\be}{\begin{equation}}
\newcommand{\ee}{\end{equation}}

\title{Heavy--light decay constants from the step scaling method\thanks{talk given at Lattice 2003}}

\author{Filippo Palombi\address[Fermi]{``E.~Fermi''~Research Center, c/o Compendio Viminale --  pal.~F, I-00184 Rome, Italy}}
       
\begin{document}

\begin{abstract}
We discuss results for the heavy--light decay constants in the continuum limit of quenched lattice QCD from finite size scaling techniques. We disentangle the 
simultaneous presence of the different energy scales characterizing heavy--light physics by first performing simulations at the unphysical volume $L_0=0.4$ fm, and 
then evolving the results towards the infinite volume. We find $f_{B_s}= 192(6)(4)$ MeV and $f_{D_s}=240(5)(5)$ MeV. The approach has been developed by
 the APE group at the University of Rome ``Tor Vergata''. 

\end{abstract}

\maketitle

\section{Introduction}
The decay constants $f_{h\ell}$ of the heavy--light pseudoscalar mesons depend upon two different energy scales. A first
energy scale is the one associated with the heavy quark mass $m_h$, and a second one is related to the light quark mass
$m_\ell$, or $\Lambda_{QCD}$. As a consequence, numerical simulations of heavy--light decay constants require tiny lattice
spacings, in order to properly describe the dynamics of the heavy quark (highly localized), and large physical volumes, to accommodate 
the light quark (widely spread). In practice, a direct reliable simulation of $f_{h\ell}$ should demand a lattice size of around $L/a\sim O(100)$,
which is hardly affordable nowadays in quenched simulations as well as in future unquenched ones. 
Different approaches are available on the market to face this problem \cite{Kronfeld}. I describe a new method which has 
been developed by the APE group at the University of Rome ``Tor Vergata'' \cite{ToV1,ToV2}, called the {\it step scaling method} at Lattice 2002 conference \cite{Yamada}.

\section{Step scaling method}
The step scaling method is a two step procedure where the heavy--light decay constants $f_{h\ell}$ are first computed on a small unphysical
volume with $L_0=0.4$ fm, and then evolved towards larger volumes in order to remove finite size effects. The evolution is simply provided by the identity
\be
f_{h\ell}(L_\infty) = f_{h\ell}(L_0)\frac{f_{h\ell}(L_1)}{f_{h\ell}(L_0)}\frac{f_{h\ell}(L_2)}{f_{h\ell}(L_1)}\dots
\ee
where $L_0<L_1<L_2\dots$ The basic ingredient of the procedure is the computation of the step scaling function
\be
\sigma(m_\ell,m_h,L_{k-1}) = \frac{f_{h\ell}(m_\ell,m_h,L_k)}{f_{h\ell}(m_\ell,m_h,L_{k-1})}\biggr|_{L_k=sL_{k-1}}
\ee
i.e. the ratio of the decay constants at two different volumes and same quark masses. We fix $s=2$ as the ratio of subsequent sizes and, after
two evolution steps, we end up on $L_2=1.6$ fm, which we take as the ``infinite'' volume, i.e. free of finite size effects. At fixed lattice spacing, both $f_{h\ell}(L_0)$ and the $\sigma$'s are simulated for a set of different quark masses, measured in a RGI quark mass scheme. Then, they are separately extrapolated to the continuum limit. Units are fixed through the $r_0$ scale \cite{r01,r02}. 
The main advantage of the method is that $f_{h\ell}(L_0)$ can be simulated at physical heavy quark masses without big lattice artifacts while the $\sigma$'s preserve in all steps a soft dependence upon the heavy quark mass, due to cancellations among volume depending terms. This makes the extrapolation to the physical values of $m_c^{RGI}$ and $m_b^{RGI}$ numerically safe. Moreover, such values can be
obtained by a self--consistent application of the step scaling method to the HL--meson masses \cite{nazario,HQM}.

\section{Computational framework}
The calculation is set up on the Schr\"odinger Functional scheme, with topological parameters
\be
T=2L,\quad C=C'=0,\quad \theta=0
\ee
$C$ and $C'$ being the gauge fields on time boundaries, and $\theta$ being a phase which affects the periodicity
of the fermion spatial boundary conditions. We simulate the basic SF correlation functions, i.e. the axial, pseudoscalar and boundary normalization bilinears  
\be
F_A^I(x_0) = -\frac{a^6}{2}\sum_{\bf yz}\langle \bar \zeta_j({\bf y})\gamma_5\zeta_i({\bf z})A_0^I(x)\rangle
\ee
\be
F_P(x_0) = -\frac{a^6}{2}\sum_{\bf yz}\langle \bar \zeta_j({\bf y})\gamma_5\zeta_i({\bf z})P(x)\rangle
\ee
\be
F_1 = -\frac{a^{12}}{3L^6}\sum_{\bf yzy'z'}\langle \bar \zeta_j({\bf y})\gamma_5\zeta_i({\bf z})\bar \zeta'_i({\bf y'})\gamma_5\zeta'_j({\bf z'})\rangle
\ee
In terms of these, a finite size definition of the decay constants is given by taking the axial correlation function in the middle of the lattice
\be
f_{h\ell} = \frac{2}{\sqrt{L^3M_A(T/2)}}Z_A(1+b_Aam)\frac{F_A^I(T/2)}{\sqrt{F_1}}
\ee
and the normalization is provided in terms of the effective meson mass, still measured in a finite size fashion
\be
aM_A(T/2) = \frac{1}{2}\ln\biggl[\frac{F_A^I(T/2-a)}{F_A^I(T/2+a)}\biggr]
\ee
Quark flavors are defined according to their RGI masses (scale/finite--size independent), defined as follows. We start from the bare quark masses
\be
am_i=\frac{1}{2}\biggl[\frac{1}{\kappa_i} - \frac{1}{\kappa_c}\biggr]
\ee
\be
m_{ij}^{WI}=\frac{(1/2)(\partial_0 + \partial_0^*)F_A + a c_A\partial_0^*\partial_0F_P}{2F_P}\biggr|_{T/2}
\ee 
and connect them to the RGI quark masses through proper renormalisation factors known in literature \cite{ZM,ZZ}
\be
m_{ij}^{RGI}=Z_M\biggl[1+(b_A-b_P)\frac{am_i+am_j}{2}\biggr]m_{ij}^{WI}
\ee
\be
\hat m_i^{RGI} = Z_MZ[1+b_mam_i]m_i, \quad Z = \frac{Z_mZ_P}{Z_A}
\ee
In this way, we get a plethora of definitions, all with the same continuum limit and differing in $O(a^2)$ terms:
\begin{itemize}
\item[1.] $m_i^{RGI} = m_{ii}^{RGI}$
\item[2.] $m_{i_{(j)}}^{RGI}= 2m_{ij}^{RGI} - m_{jj}^{RGI}, j\ne i$
\item[3.] $\hat m_i^{RGI}$
\end{itemize}
After choosing a target pair $(m_i^{RGI},m_j^{RGI})$, we tune the hopping parameters $(\kappa_i,\kappa_j)$ such that the values of the masses
coming from the various definitions equate the chosen pair. Of course, each choice of $(\kappa_i,\kappa_j)$ produce a different value of $f_{h\ell}$, 
all the values differing in $O(a^2)$ terms, as it can be seen in Fig.1.
\begin{figure}[h]
\vskip -0.7cm
\includegraphics[width=7cm]{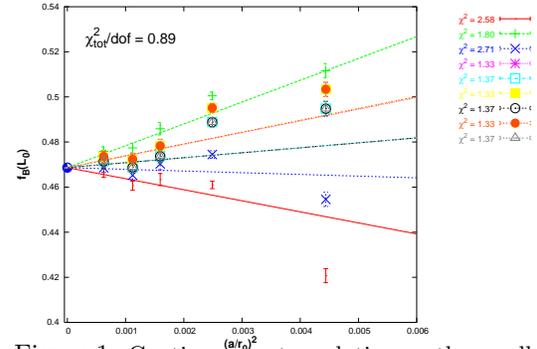}
\vskip -1.2cm
\caption{\small Continuum extrapolation on the small volume of $f_{B_s}(L_0)$. Units in GeV.}
\label{fig:6737}
\vskip -1.0cm
\end{figure}
\section{Results at $L_0=0.4$ fm}
Simulations of the decay constants on the smallest volume $L_0=0.4$ fm have been performed at five different
lattice spacings, using the geometries $24\times 12^3$, $32\times 16^3$, $40\times 20^3$, $48\times 24^3$ and
$64\times 32^3$. For each lattice spacing a set of eight RGI quark masses have been simulated, two around
$m_b^{RGI}$, two around $m_c^{RGI}$, one additional at 4.0 GeV, and three light quark masses at 140, 100 and
60 MeV. We fix the light quark mass at the physical value of $m_s^{RGI}$ by interpolation and we get 
\be
f_{B_s}(L_0) = 475(2){\rm MeV}
\ee
\be
f_{D_s}(L_0)=644(3){\rm MeV}
\ee
The errors quoted at this stage are statistical only, and they have been evaluated by a jackknife procedure.
The continuum extrapolation has been obtained by performing a combined linear fit in $(a/r_0)^2$ over the last
three sets of points of Fig.1. We estimate a systematic error of 1\%, related to this extrapolation, by the discrepancy observed 
after restricting the fit to the two sets of points nearest to the continuum.
The results obtained are clearly unphysical, as they have to be corrected through the evolution factors given by the $\sigma$'s.
\section{Evolution steps}
The first evolution step connects $L_0$ to $L_1=2L_0=0.8$ fm. Three different lattice spacings have been simulated
in order to perform the continuum extrapolation, corresponding to the geometries with $L_0/a=8,12,16$ and consequently $L_1/a=16,24,32$.
The values of the simulated quark masses have been halved in order to leave the lattice artifacts unchanged. In Fig.~2 the continuum step
scaling function $\sigma(L_0)$ is plotted vs. the inverse of the heavy quark mass with the light quark mass fixed at $m_s^{RGI}$. 
\begin{figure}[h]
\vskip -0.8cm
\includegraphics[width=7cm]{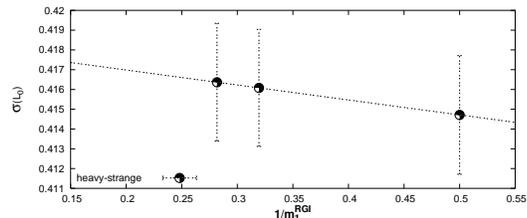}
\vskip -1.1cm
\caption{\small Continuum step scaling function $\sigma(L_0)$ as a function of $1/m_h^{RGI}$ at $m_\ell^{RGI} = m_s^{RGI}$. Units in GeV.}
\label{fig:6737}
\vskip -0.7cm
\end{figure}
At this stage we get
\be
\sigma_{B_s}(L_0) = 0.417(3)
\ee
\be
\sigma_{D_s}(L_0) = 0.414(3)
\ee
Analogously, the second evolution step connects $L_1$ to $L_2=2L_1=1.6$ fm. Also here, three lattice spacings have been simulated with the
same lattice geometries as before. Again the values of the simulated quark masses have been halved. Fig.~3 shows the continuum step scaling
function $\sigma(L_1)$ vs. the inverse of the heavy quark mass at $m_l^{RGI}=m_s^{RGI}$. We find
\be
\sigma_{B_s}(L_1) = 0.97(3)
\ee
\be
\sigma_{D_s}(L_1) = 0.90(2)
\ee
\begin{figure}[h]
\vskip -1.0cm
\includegraphics[width=7cm]{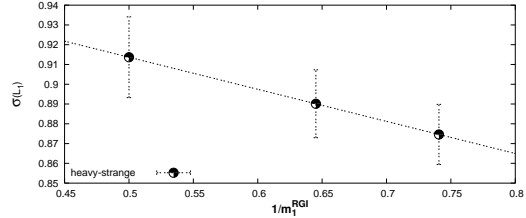}
\vskip -0.9cm
\caption{\small Continuum step scaling function $\sigma(L_1)$ as a function of $1/m_h^{RGI}$ at $m_\ell^{RGI} = m_s^{RGI}$. Units in GeV.}
\label{fig:6737}
\vskip -1.0cm
\end{figure}
\section{Results}
Combining the results of the small volume $L_0$ with the evolution factors, we obtain
\be
f_{B_s} = 192(6)(4){\rm \ MeV}
\ee
\be
f_{D_s} = 240(5)(5){\rm \ MeV}
\ee
The first error is statistical while the second one is systematic and due to the uncertainties on the continuum extrapolation, on the scale
and on the renormalisation factors.



\end{document}